\documentclass[12pt]{article}
\usepackage{amsmath}
\usepackage{amssymb}
\numberwithin{equation}{section}

\newcommand{\reff}[1]{{\rm (\ref{#1})}}
\newcommand{\Proof}{{\em Proof:} }
\def\inv{{^{-1}}}
\newtheorem{theorem}{Theorem}[section]
\def\qed{\ {\it q.e.d.}}
\def\o{\omega}
\begin{document}

\title{Acoustic Scattering and the Extended Korteweg deVries 
Hierarchy}

\author{R. Beals\thanks{Research supported by NSF grant DMD-9423746}\\
Yale University \\New Haven, CT 06520
\and
D. H. Sattinger\thanks{Research supported by NSF Grant DMS-9501233.}\\
University of Minnesota\\Minneapolis 55455
\and
J. Szmigielski\thanks{Research supported by the Natural Sciences and 
Engineering Research Council of Canada}\\
University of Saskatchewan\\
Saskatoon, Saskatchewan, Canada}

\maketitle

\centerline{{\it Advances in Mathematics}, {\bf 140} (1998), 190-206}

{\abstract The acoustic scattering operator on the real line is 
mapped to a Schr\"odinger operator under the Liouville transformation. The 
potentials in the image are characterized 
precisely in terms of their scattering data, and the inverse transformation is
obtained as a simple, linear quadrature. An existence theorem for the 
associated Harry Dym flows is proved, using the scattering method. 
The scattering problem associated with the Camassa-Holm flows on the real line 
is solved explicitly for a special case, which is used to reduce a general
class of such problems to scattering problems on finite intervals.}


\section{Introduction}

In this paper we consider the forward and inverse scattering problems on the 
line for operators 
\begin{equation}\label{scat}
L_k=D^2+k^2\rho^2-q, \qquad D=d/dx.
\end{equation}
For $\rho=1$, \reff{scat} is the Schr\"odinger operator and $k$ is the wave
number, or momentum.  As is well-known, the 
scattering data of \reff{scat} evolves linearly as $q$ evolves according to 
the nonlinear Korteweg-deVries flows. 

For $q=0$ and $\rho\to 1$ as $x\to\pm\infty$ \reff{scat} is the {\it acoustic} 
scattering problem. In this interpretation, $k$ is the frequency. 
The associated hierarchy of flows, of which the first is
\begin{equation}\label{harry}
(\rho^2)_t={1\over 2}D^3\Big({1\over\rho}\Big),
\end{equation}
were introduced by Martin Kruskal \cite{kruskal} and attributed
to an unpublished paper of Harry Dym. 

For $q=1/4$ and $\rho\to 0$ at infinity, \reff{scat} was introduced by Camassa 
and Holm \cite{ch}, \cite{chh}, in connection with a nonlinear shallow water 
model.  The Camassa-Holm equation itself\footnote{The equation itself had been 
obtained previously by {\sc Fuchsteiner} \cite{fuch} by the method of 
recursion operators; but that method does not give the isospectral operator.}, 
in the normalization determined by  \reff{scat} with $q=1$, is
\begin{gather}
 (1-\frac14 D^2)u_t=\frac32 (u^2)_x-\frac18 (u_x^2)_x-\frac14 (uu_{xx})_x, \label{cholm} \\
\rho^2=2(1-\frac14 D^2)u. \nonumber
\end{gather}
 
For this equation, considered on the line, it is natural to assume that 
the potential $\rho\to 0$ as $|x|\to \infty$; 
the scattering problem on the line is therefore singular.
The case of periodic potentials has been resolved by Constantin and McKean 
\cite{cmck}. 
Camassa and Holm \cite{ch}, and Camassa, Holm and Hyman \cite{chh} have also obtained
Blow-up results for certain initial data and have constructed two-soliton
solutions by direct methods -- that is, methods not based on the inverse
scattering technique. Constantin and Escher \cite{ce}  have proven global existence theorems for the
Camassa Holm equation on the real line for a large class of initial data.

The scattering problem for \reff{scat} when $\rho\equiv 1$ is by now well 
studied and completely understood; the inverse problem can be solved by the 
integral equation method of Gel'fand and Levitan \cite{gl}. In the acoustic 
problem, $\rho\to 1$ as $|x|\to\infty$ and $q=0$, it is well-known that
the classical Liouville transformation takes $L_k$
to the Schr\"odinger operator \cite{bu}.   
The inverse transformation from the Schr\"odinger to the acoustic problem 
requires the inversion of a differential equation. 
We show in this paper that %by a simple, yet standard device 
this step can be reduced to a simple linear quadrature. 
This greatly simplifies the inversion problem.

It is the acoustic problem, rather than the Schr\"odinger equation, that is 
relevant to physical applications in elasticity and reflection seismology 
\cite{ba1}, \cite{ba3}, \cite{ba2}, \cite{bu}, \cite{symes}.
\footnote{We thank {\sc Fadil Santosa}, of the University of Minnesota, for 
his helpful information regarding the applied literature in acoustic 
scattering problems.}

Although the scattering problem for \reff{scat} when $\rho\to 0$ at infinity 
and $q\equiv 1$ is
{\it a priori} singular, a particular change of variables reduces 
it to the Dirichlet problem
$$
\left(\frac{d^2}{d\xi^2}+k^2\rho_1^2(\xi)\right)\psi(\xi,k)=0,
\qquad
-1\le \xi \le 1;
\qquad
\psi(\pm 1,k)=0.
$$
This is a density problem on a finite interval \cite{krein1}, \cite{krein2}. 
A second Liouville transformation converts this to a Schr\"odinger 
problem on a finite interval.  Thus the inversion problem reduces to 
solving an inverse Sturm-Liouville problem on a finite interval 
\cite{jl}, \cite{pt}.

\section{Acoustic Scattering}

The equation
\begin{equation}\label{ac}
(D^2+k^2\rho^2)\psi=0,
\qquad
-\infty < x < \infty,
\end{equation}
arises in scattering problems for the wave equation
$$
u_{tt}-c^2(x)u_{xx}=0,
$$
where the sound speed $c$ is $1/\rho$.  Therefore it is natural to assume
$\rho$ to be bounded away from $0$ and to have a finite limit
as $|x|\to\infty$.  We call \reff{ac} the acoustic scattering problem.
We assume throughout that $\rho$ is real and positive, while $\rho-1$
belongs to ${\cal S}$, the Schwartz class of rapidly decaying functions. 

We begin by constructing the wave functions of \reff{ac}, normalized at 
$\pm \infty$, by the WKB method. We write the wave functions in the form
$$
\varphi_+(x,k)=\ell_+(x,k)e^{-ikS(x)},
\qquad
\psi_+(x,k)=m_+(x,k)e^{ik(S(x)},
$$
where $\ell_+$ and $m_+$ are normalized to be $1$ at $-\infty$ and $+\infty$ 
respectively. 
Substituting this form of the wave functions into equation
\reff{ac}, we find that $m_+$ satisfies the differential equation
$$
m''+2ikS'm'+[ikS''+k^2(\rho^2-S^{\prime 2})]m=0.
$$
The term involving $k^2$ is eliminated by requiring $S(x)$ to satisfy the 
{\it eiconal equation}
$$
(S')^2=\rho^2.
$$
We take 
$$
S'(x)=\rho(x), 
\qquad
S(x)=x+\int_{-\infty}^x [\rho(y)-1]\,dy.
$$
Then
$$
S(x)= \begin{cases} x+o(1) & x\to -\infty\\ x+\gamma +o(1) & x\to\infty \end{cases}
$$
where
$$
\gamma=\int_{-\infty}^\infty[\rho(y)-1]\,dy.
$$

We now have
\begin{equation}\label{m}
m''+2ik\rho m'+ik\rho' m=0,
\qquad
m \to 1\ \ \mbox{as}\ \ x\to\infty. 
\end{equation}
The solution to this differential equation is constructed by converting 
it to the Volterra integral equation
\begin{align}
m(x,k)=&1+\int_x^\infty {e^{2ik(y-x)} -1 \over 2ik} (2ik(1-\rho)m'-ik\rho' m)\,dy 
\nonumber \\[4mm]
=&1+\int_x^\infty G(x,y,k) m(y,k)\,dy, \label{volterra}
\end{align}
where
$$
G(x,y,k)=2ike^{2ik(y-x)}(\rho(y)-1)+\frac12(e^{2ik(y-x)}-1)\rho'(y).
$$
This integral equation can be solved for any $k$ in the upper half
plane $\Im k\ge 0$ by the method of 
successive approximations, since $\rho-1$ and $\rho'$ belong to 
${\cal L}^1(\mathbb R)$.  The solution is analytic with respect to $k$; we denote it by
$m_+(x,k)$.  We denote the Schwarz reflection of $m_+$ to the lower half 
plane by $m_-(x,k)=\overline{m_+(x,\bar k)}$, and extend $\psi_+$ to the 
lower half plane accordingly:
$$
\psi_-(x,k)=\overline{\psi}_+(x,\bar k)=m_-(x,k)e^{-ikS(x)}.
$$

Similarly the function $\ell_-$ defined by
\begin{equation}\label{ell}
\ell_-(x,k)=1-\int_{-\infty}^x G(x,y,k)\ell_-(y,k)\,dy.
\end{equation}
is analytic with respect to $k$ in the lower half plane and
has a Schwarz reflection $\ell_+(x,k)$ analytic in the upper half plane.
We denote the corresponding wave functions by
by
$$
\varphi_\pm(x,k)=\ell_\pm(x,k)e^{\mp ikS(x)}.
$$

The boundary values on the real line satisfy the asymptotic conditions
\begin{equation}\label{norm}
\lim_{x\to -\infty}e^{\pm ikS(x)}\varphi_\pm=
\lim_{x\to\infty}e^{\mp ikS(x)}\psi_\pm=1.
\end{equation}
Note that 
\begin{equation}\label{specialk}
\ell_+(x,k)=\ell_-(x,-k),\qquad m_-(x,k)=m_+(x,-k),\qquad k\in\mathbb R.
\end{equation}
Moreover $G(x,y,0)=0$ so
\begin{equation}\label{k0}
\ell_\pm(x,0)=m_\pm(x,0)\equiv 1.
\end{equation}

The {\it scattering data\/} for the acoustic equation is
defined just as it is for the Schr\"odinger equation. 
For real non-zero $k$ each pair of wave functions $\psi_\pm$, $\varphi_\pm$ 
is (generically) independent and thus constitutes a 
fundamental set of solutions. Therefore
\begin{equation}\label{conn}
\varphi_+(x,k)=a(k)\psi_-(x,k)+b(k)\psi_+(x,k)
\end{equation}
for some functions $a$ and $b$.  For real $k$ we use \reff{specialk},
the definitions of the wave functions, and the limits, to see that \reff{conn}
becomes
$$
\ell_+(x,k)\sim a(k)+b(k)e^{2ikS(x)}\sim a(k)+b(k)e^{2ik(x+\gamma)}
\quad{\mbox{ as}}\ x\to+\infty.
$$
In view of \reff{ell}, \reff{specialk} and the form of $G$, we obtain for real $k$ that
\begin{align}\label{ab}
a(k)&=1+{1\over2}\int^\infty_{-\infty}\rho'(y)\ell_+(y,k)\,dy;\\
b(k)&=\int^{\infty}_{-\infty}e^{-2ik(y+\gamma)}
\big\{2ik(\rho(y)-1)-\tfrac12\rho'(y)
\big\}\ell_+(y,k)\,dy.\nonumber
\end{align}
It follows from \reff{ab} and \reff{k0} that
\begin{equation}\label{ab0}
a(0)=1, \qquad b(0)=0.
\end{equation}

The formula for $a(k)$ in \reff{ab} extends to the upper half
plane.  We take the Wronskian of both sides of \reff{conn} with
$\psi_\pm$ and obtain
\begin{equation}\label{wronsk}
a(k)=\frac{W(\varphi_+,\psi_+)}{W(\psi_+,\psi_-)}
=\frac{W(\varphi_+,\psi_+)}{2ik},\quad 0\ne k\in\mathbb R.
\end{equation}
In fact, taking asymptotics as $x\to\infty$,
$$
W(\psi_-,\psi_+)=W(e^{-ikS},e^{ikS})=2ikS'\sim 2ik=2ik\rho\sim 2ik.
$$
The expression \reff{wronsk} for $a(k)$ also extends to the upper
half plane.
\medskip
\begin{theorem}\label{a} The reduced wave functions $m_+,\ \ell_+$ are analytic in the upper half plane, 
and $m_-,\ \ell_-$ are analytic in the lower half plane.

The function $a(k)$ is analytic in the upper half plane.  
Moreover $a(0)=1$ and $a$ has no zeros. 
\end{theorem}

\Proof We have proved everything but the last statement.
As in the case of the Schr\"odinger equation, the acoustic scattering 
data for real $k$ satisfy
\begin{equation}\label{symm}
a(k)=\overline{a(-k)},\quad b(k)=\overline{b(-k)};\qquad
|a(k)|^2-|b(k)|^2=1.
\end{equation}
Therefore $a$ has no real zeros.  According to \reff{wronsk} a
zero at $k=i\o$ in the upper half plane would 
correspond precisely to a {\it bound state\/}: an
$L^2$ wave function with negative energy.  This would be an eigenfunction for the
operator $L=\rho^{-2}D^2$ with eigenvalue $-k^2=\o^2>0$.  However $L$
is selfadjoint and negative in $L^2(\mathbb R,\rho^2\,dx)$, so it cannot
have such an eigenvalue.\  \ \ \qed 

\medskip
We define the reflection coefficient $r$ by
$$
r(k)=\frac{b(k)}{a(k)}, 
\qquad
\Im k=0.
$$
In the absence of bound states, $r$ constitutes the complete scattering data 
for the problem.  In fact from \reff{symm} we have
$$
|a|^2=\frac{1}{1-|r|^2}.
$$
Since $a$ is analytic in $\Im\,k\ne 0$, tends to 1 as $k\to\infty$, and has no 
zeros, $\arg\,a$ can be recovered from $\log|a|$ on the real axis by the 
Hilbert transform:
$$
\arg\,a(k)=\frac{1}{\pi}{\rm P.V.}\int_{-\infty}^\infty \frac{\log\,|a(t)|}{k-t}dt.
$$
Then $\log\,a$ is obtained for $\Im\,k\ne 0$ by the Cauchy integral 
representation.

\section{The Liouville Transformation}

The acoustic equation may be transformed to the Schr\"odinger equation by the 
well-known Liouville transformation.  By our assumptions, $S(x)$ is a monotone 
increasing function on the line, hence we may define a change of variables by 
$\xi=S(x)$. The variable $\xi$ corresponds physically to the time of travel.  
By the chain rule,
$$
\frac{d}{dx}=\frac{d\xi}{dx}\frac{d}{d\xi}=\rho\frac{d}{d\xi},
\qquad
\frac{d^2}{dx^2}=\rho\frac{d}{d\xi}\rho\frac{d}{d\xi}.
$$
To keep track of the relevant variables we define $\rho_s(\xi)$ by
$$
\rho_s(\xi)=\rho(x),\quad\mbox{at}\ \xi=S(x).
$$
The mapping $f\to f\circ S\inv$ is a unitary map from $L^2(\mathbb R,\rho^2\,dx)$
to $L^2(\mathbb R,\rho_s\,d\xi)$.  It carries the negative selfadjoint operator
$\rho^{-2}D^2$ to
the operator
$$
D_\xi^2+\frac{D_\xi\rho_s}{\rho_s}D_\xi, \qquad D_\xi={d\over d\xi}.
$$
To complete the Liouville transformation we use the unitary
map $f\to f\sqrt{\rho_s}$ from $L^2(\mathbb R,\rho_s\,d\xi)$ to $L^2(\mathbb R, d\xi)$.
The corresponding gauge transformation (conjugation by ${\rho_s}^{-1/2}$)
takes the preceding operator to the Schr\"odinger operator
\begin{equation}\label{schrod}
D_\xi^2-q(\xi),
\end{equation}
where
\begin{equation}\label{riccati}
q(\xi)=\frac12\frac{D_\xi^2\rho_s}{\rho_s}-
\frac14\Big(\frac{D_\xi\rho_s}{\rho_s}\Big)^2.
\end{equation}
Since $D_\xi=\rho^{-1}D_x=\rho\inv D$, we can also express the potential as
\begin{equation}\label{schwarzian}
q(S(x))={1\over 2}{D^2\rho\over\rho^3}-{3\over4}\Big({D\rho\over\rho^2}\Big)^2
=\frac{1}{\rho^2}\left( -\sqrt\rho D^2\left(\frac{1}{\sqrt\rho}\right)\right).
\end{equation}

We remark that 
$$
q(S(x))={1\over 2\rho^2}\{\xi,x\}
$$
where $\{\xi,x\}$ denotes the Schwarzian derivative
$$
\{\xi,x\}=\frac{D^3\xi}{D\xi}-\frac32\left(\frac{D^2\xi}{D\xi}\right)^2.
$$

We have seen that the transformation from  the Schr\"odinger potential 
$q$ to the associated  
acoustic potential $\rho$ is given by the equation \reff{riccati},
together with a change of variables.  We show below that this
transformation may be computed by a linear operation on
scattering data and a simple quadrature.
\medskip
\begin{theorem}\label{invariant} The normalized wave functions $\varphi_{s,\pm}$,
$\psi_{s,\pm}$ for the operator
\reff{schrod},\reff{riccati} are related to the wave functions for the
acoustic operator by
\begin{equation}\label{schwave}
\varphi_{s,\pm}(\xi)=\sqrt{\rho(\xi)}\,\varphi_\pm(S\inv\xi);
\quad \psi_{s,\pm}=\sqrt{\rho(\xi)}\,\psi_\pm(S\inv\xi).
\end{equation}
The scattering data $(a,b)$ of the Schr\"odinger problem is 
precisely that obtained for the acoustic problem \reff{ac}.
\end{theorem}
\Proof  The relation in \reff{schwave} simply implements the two
unitary transformations, change of variables and change of gauge.
Therefore functions $\varphi_{s,\pm}$, $\psi_{s,\pm}$ are
wave functions for the operator \reff{schrod}.  Moreover they have
the correct asymptotics and analyticity properties.\qquad$\qed$

\medskip
\begin{theorem}\label{range} The image of the acoustic problem under the Liouville 
transformation consists of all Schr\"odinger operators with Schwartz
class potentials and no bound states, such that $a(0)=1$.

The inversion of the Liouville transformation is given by
\begin{equation}\label{inversion}
\rho_s(\xi)=\psi_s(\xi,0)^2,
\qquad
x=\xi+
\int_{-\infty}^\xi \left(\frac{1}{\rho_s(\xi')}-1\right)\,d\xi'.
\end{equation}
\end{theorem}
\Proof Theorems \ref{a} and \ref{invariant} imply that operators
in the range of the Liouville transformation have no bound states and
have $a(0)=1$.  The transformation equation \reff{riccati} shows
that the potential is of Schwartz class.
The acoustic potential $\rho$ may be recovered from the 
Schr\"odinger potential $q$ as follows. By \reff{schwave} the wave functions 
at $k=0$ are related by
$$
\psi_s(\xi,0)=\sqrt{\rho_s(\xi)}\,\psi(x,0).
$$
On the other hand we have observed that the normalized
acoustic wave functions at $k=0$ are identically $1$.
It follows that
$$
\psi_s(\xi,0)=\sqrt{\rho_s(\xi)}.
$$
Therefore to reconstruct the function $x=x(\xi)$ from the
Schr\"odinger potential, we can compute $\psi_s(\cdot,0)$ 
and use
$$
\frac{dx}{d\xi}=\frac{1}{\rho_s(\xi)}=\frac{1}{\psi_s(\xi,0)^2}.
$$
 
Suppose, conversely, that $q$ is a Schwartz class Schr\"odinger
potential with no bound states, such that $a(0)=1$.  Then $\psi_s(\cdot,0)
=\varphi_s(\cdot,0)$  is real and asymptotically $1$ in each
direction.  We prove that $\psi_s(\xi,0)=m^s(\xi,0)$ has no zeros when there 
are no bound states. First, note that $m^s_+(\cdot,i\o)$ is real 
and converges uniformly to $m^s_+(\cdot,0)$ as $\o$ 
decreases to $0$.  If the latter
function had any zeros they would necessarily be simple, since
$m^s$ is a solution of a second order differential equation,
and therefore $m^s_+(\cdot,i\o)$ would have zeros
for small $\o>0$.
This function converges uniformly to $1$ as $\varepsilon\to
+\infty$ and the zeros remain simple, so there would be a 
value $\o>0$ for which $m^s_+$ has limit $0$ as $x\to\infty$.
But the corresponding $\psi^s_+$ would be a bound 
state.  Therefore $\psi_s(\cdot,0)$ is positive.
\footnote{By the same argument one can prove that the number of 
zeros of $m^s(\xi,0)$ is precisely equal to the number of bound states. 
If there are $n$ bound states, then by oscillation theory $m^s(\xi,i\o_n)$ 
has $n$ zeros. The number of zeros of 
$m^s$ is constant as $\o$ decreases from $\o_n$ to zero; so $m^s(\xi,0)$ 
also has $n$ zeros.} 
Because of this we can use
\reff{inversion} to construct a change of variables and a 
potential $\rho$.  The assumption that $q$ belongs to $\cal S$
implies that $\psi_s(\cdot)-1$ belongs to $\cal S$.  Therefore
$\rho-1$ belongs to $\cal S$.  Clearly $D^2-q$ is the Liouville transform
of the acoustic operator associated to $\rho$.  \qquad$\qed$

\medskip

We have shown that the density $\rho$ in the time of travel coordinate
$\xi$ is obtained immediately from the Schr\"odinger wave function at $k=0$.
The latter can be obtained directly from the Gel'fand Levitan kernel 
$K(x,y)$, since the wave functions are given by
$$
\psi(\xi,k)=e^{ik\xi}+\int_\xi^\infty K(\xi,y)e^{iky}\,dy.
$$

\section{The Harry Dym Flows}

The Harry Dym flows are related to the operator $L=\rho^{-2}D^2$ 
exactly as the KdV flows are to  $D_\xi^2-q$.  They are derived 
from commutator conditions $[L,A]=0$ where $A$ is the Liouville 
transformation of an operator that determines one of the 
KdV flows.  The computation of $A$ is rather complicated, and seems
to shed no light on the issue.  It is therefore more convenient to 
work with $L_k=\rho^2(L+k^2)$ instead.  Then the commutator condition must 
be modified to 
\begin{equation}\label{lax}
[L_k,A]=B\,L_k
\end{equation}
where $L_k,\, A,\, B$ act on functions $\psi=\psi(x,t,k)$, and $L_k$
and $A$ are given by:
$$
L_k=D^2+k^2\rho^2,\qquad A=-{\partial\over\partial t}+\sum^n_{j=0}
f_jD^j.
$$
The coefficients $f_j$ are taken to be polynomials in $k^2$.  The 
significance of \reff{lax} is that $[L_,, A]$ vanishes on the
wave functions, i.e. solutions of $L_k\psi=0$.  The conditions relating 
the coefficients of $L_k$ and $A$ may be 
obtained from cross-differentiation of the pair of equations
$L_k\psi=0$, $A\psi=0$.
  
Any operator product of the form $CL$ may be added to $A$
without affecting \reff{lax}.
Therefore even powers of $D$ may be eliminated in favor of even powers
of $k$, and it is enough to take $n=1$: $A=-\partial_t+fD+g$.   Then
\begin{align*}
[L,A]
=&2f_xD^2+(f_{xx}+2g_x)D+[k^2(\rho^2)_t
-k^2f(\rho^2)_x]+g_{xx}\\[4mm]
=&(f_{xx}+2g_0)D+[k^2(\rho^2)_t
 -k^2f(\rho^2)_x+g_{xx}-2k^2\rho^2f_x]+2f_1D\,L.
\end{align*}
The coefficient of $D$ must vanish, so we take 
$$
g=-\frac12 f_x.
$$
Then the conditions for \reff{lax} become
\begin{align*}
k^2(\rho^2)_t&={1\over 2}\,f_{xxx}+k^2\big[(\rho^2)x f+2\rho^2 f_x\big]
\\[4mm]
&={1\over2}\,f_{xxx}+2\rho(\rho f)_x.\end{align*}
Setting
$$
f=\sum^n_{j=1}F_jk^{2j},
$$
substituting this expession into the previous identity, and comparing 
coefficients of powers of $k$, we find that $F_n$ is a constant multiple 
of $1/\rho$, $F_0=0$, and the remaining coefficients
can be determined from the recursion relation
$$
4\rho D(\rho F_{j-1})=-D^3F_j.
$$
The flow of $\rho$ is
$$
(\rho^2)_t={1\over 2}D^3F_1.
$$
For $n=1$, we take $F_1=1/\rho$ and obtain \reff{harry}.

\medskip
\begin{theorem}\label{evolve} Under the Harry Dym flow, the scattering data $a(k,t)$ and 
$b(k,t)$ evolve according to
\begin{equation}\label{sdevol}
\dot a=0, \qquad
\dot b=2ik^3b.
\end{equation}
\end{theorem}
\Proof The commutator relation \reff{lax} implies that the kernel of $L$ 
is invariant under $A$; that is, 
$$
L\varphi(x,t,k)=0 \quad \Longrightarrow\quad LA\varphi(x,t,k)=0.
$$

The wave function 
$$
\varphi_+(x,t,k)=\ell_+(x,t,k)e^{-ikS(x)}\sim e^{-ikx},
\qquad
x\to-\infty.
$$
On the other hand, for the Dym equation itself, 
$$
A=-\partial_t+\frac{k^2}{\rho}D-\frac12 D\left(\frac1\rho \right)\sim -\partial_t+k^2D,
\qquad
x\to \pm\infty.
$$
Therefore,
$$
A\varphi_+(x,t,k)\sim (-\partial_t+k^2D)e^{-ikx}=-ik^3 e^{-ikx},
\qquad
x\to -\infty.
$$
Since the wave functions are uniquely determined by their asymptotic 
behavior as $x\to-\infty$, we conclude that
$$
A\varphi_+=-ik^3\varphi_+.
$$

Similarly, we find that
$$
A\psi_\pm=\pm ik^3\psi_\pm,
\qquad
A\varphi_-=ik^3\varphi_-.
$$
Therefore, applying $A$ to \reff{conn}, we obtain
\begin{align*}
A\varphi_+=&-ik^3\varphi_+=-ik^3(a\psi_-+b\psi_+)\\
=&-\dot a \psi_--\dot b\psi_++aA\psi_-+bA\psi_+\\
=&-\dot a \psi_--\dot b\psi_+-ik^3a\psi_-+ik^3b\psi_+;
\end{align*}
and equations \reff{sdevol} follow immediately from the independence of
$\psi_+$ and $\psi_-$.   $\qed$

\section{The Camassa-Holm spectral problem}

We consider the operators
\begin{equation}\label{chiso}
L=D^2+k^2\rho^2-1,
\end{equation}
where $D=d/dx$, introduced by Camassa and Holm \cite{ch}. We are 
interested in a singular case of \reff{chiso}:
$\rho$ is positive but $\rho\to 0$ rapidly  
at infinity.  As a consequence, the natural normalization for the
wave functions is independent of $k$:
\begin{equation}\label{chwave}
\lim_{x\to-\infty}e^{-x}\varphi(x)=1=\lim_{x\to+\infty}e^x\psi(x).
\end{equation}

The Camassa-Holm equation \reff{cholm} implies that the evolution of
$\rho^2$ is given by
$$
(\rho^2)_t=u(\rho^2)_x+2u_x\rho^2.
$$
Under our assumption that $\rho$ is strictly positive, we have the
equivalent form
\begin{equation}\label{cholm2}
\rho_t=(u\rho)_x,\qquad (4-D^2)u=2\rho^2.
\end{equation}
Let us assume that
\begin{equation}\label{int}
\int^\infty_{-\infty}e^{2|y|}\rho(y)^2\,dy<\infty.
\end{equation}
Then, from the second equation in \reff{cholm2}, 
$$
u(x)={1\over2}\int^\infty_{-\infty}e^{-2|x-y|}\rho(y)^2\,dy.
$$
It follows that 
\begin{equation}\label{asexp1}
\lim_{x\to\pm\infty}e^{2|x|}u(x)=\int^\infty_{-\infty}e^{\pm2y}\rho(y)^2\,dy.
\end{equation}
Therefore the evolution
\reff{cholm2} is consistent with the assumption \reff{int},
and with the stronger assumption $\rho(x)=O(e^{-2|x|})$
as $|x|\to\infty$.  

For reasons that will become clear, a natural
class of functions $\rho$ to consider are positive $C^\infty$
functions that have asymptotic expansions
\begin{equation}\label{asexp2}
\rho(x)\sim\sum^\infty_{\nu=1}a^\pm_\nu e^{-2\nu |x|}\quad\mbox{as}
\ x\to\pm\infty,\quad a^-_1=a^+_1>0.
\end{equation}
It follows that $u$ has similar expansions.  The leading terms $b^\pm_1$
in these expansions are given by \reff{asexp1} and the remaining terms can be computed from \reff{asexp2} and the second equation in \reff{cholm2}.
These expansions are consistent with the evolution \reff{cholm2}
and the time dependence of the coefficients can be computed from
these equations.

We observe next that for a particular choice of potential $\rho$ the
Liouville transformation trivializes the operator $\rho^{-2}(D^2-1)$.
By \reff{schwarzian}, the latter transforms to
a Schr\"odinger operator with potential
\begin{align}
q(S(x))&=\frac1{\rho^2}\Big[\frac12{D^2\rho\over\rho}
-\frac34\Big({D\rho\over\rho}\Big)^2+1\Big]\nonumber\\[4mm]
&=\frac{1}{\rho^2}\Big[-\sqrt{\rho}\,D^2\Big({1\over\sqrt\rho}\Big)+1\Big].
\nonumber
\end{align}
Therefore if we choose
\begin{equation}\label{vacuum}
\rho_0(x)={1\over \cosh^2 x}=\text{sech}^2 x 
\end{equation}
the potential $q_0$ vanishes.  Note that $\rho_0(x)$ has an expansion
\reff{asexp2}.  Since $\text{sech}^2 x=D\,\tanh\, x$,
we can take as transformed variable $\zeta=S_0(x)=\tanh\,x$.  
Then the range of the transformation is the
interval $-1<\zeta<1$.\footnote{It was already remarked by Camassa and Holm  \cite{ch} that the spectrum of the operator \reff{chiso} is discrete when $\rho\to 0$ sufficiently rapidly as $x\to\pm\infty$; and they analyzed the special case when
$\rho_0^2=A\text{sech}^2(x)$. Note, however, that it is the potential $\rho_0^2=\text{sech}^4(x)$ that trivializes under the Liouville transformation.}

The positive smooth potentials $\rho$ that have asymptotic
expansions \reff{asexp2} are precisely the potentials that can be
written in the form
\begin{gather}
\rho(x)=g(\tanh\,x)\,\rho_0(x)={g(\tanh\,x)\text{sech}^2 x},\\[4mm]
g\in C^\infty([-1,1]),\quad g>0, \quad g(-1)=g(1).
\end{gather}
The Liouville transformation generated by $\rho_0(x)$ brings 
$\rho_0^{-2}(D^2+k^2\rho^2-1)$ to the form
\begin{equation}\label{gstring}
\Big({d\over d\zeta}\Big)^2+k^2g(\zeta)^2.
\end{equation}
This is the density problem for a string on the finite interval 
\cite{krein1}, \cite{krein2}.

The corresponding wave functions are related by
\begin{equation}\label{bc}
\psi_s(\zeta,k)=\sqrt{\rho_s(\zeta)}\psi_a(x,k)=
\sqrt{\rho(x)}\psi_a(x,k)
=\text{sech}\,x \psi_a(x,k).
\end{equation}
Hence the corresponding eigenfunctions vanish as $\zeta\to \pm 1$, and the appropriate boundary conditions for \reff{gstring} are Dirichlet boundary conditions.

The Liouville transformation from \reff{chiso} with $d\xi/dx=\rho$
and the Liouville transformation from \reff{gstring} with $d\xi/d\zeta
=g(\zeta)$ arrive at the same point: a Schr\"odinger
operator on a finite interval.  The length of the interval is
$2M$ where
\begin{equation}\label{interval}
M={1\over2}\int^\infty_{-\infty}\rho(x)\,dx=
{1\over2}\int^1_{-1} g(\zeta)\,d\zeta.
\end{equation}
and we normalize the interval to be $(-M,M)$.
The expression for the potential $q(\xi)$ in terms of $g$
is given by \reff{schwarzian}:
\begin{equation}\label{riccati2}
q(\xi(\zeta))=\frac1{2g^2}\{\xi,\zeta\}.
\end{equation}

\section{Camassa-Holm scattering data}

In this section we describe the scattering data for \reff{chiso}
and find its evolution 
under the Camassa-Holm flows.

We have seen that the Liouville transformation is a unitary equivalence 
between the negative operator
$\rho^{-2}(D^2-1)$ and the Schr\"odinger problem on a finite interval.
Under this transformation the normalized wave functions 
\reff{chwave} are multiplied
by $\rho$, which vanishes at $\infty$.  On the finite interval, the
asymptotic conditions become Dirichlet conditions: vanishing at an
endpoint.  Thus it is natural to take
as scattering data for \reff{chiso} the Dirichlet spectrum and the
associated coupling coefficients.  Note that the wave functions for
\reff{gstring} at $k=0$ have the form $a\zeta+b$; the corresponding
wave functions for the Schr\"odinger operator do not satisfy Dirichlet
conditions.  Thus the Dirichlet eigenvalues are strictly negative. 

At the eigenvalues $-k_n^2$ the wave functions
$\varphi$ and $\psi$ are linearly dependent:
there is a {\it coupling coefficient\/} $c_n$ such that
$$
\varphi(x,k_n)=c_n\psi(x,k_n).
$$
Thus we take as scattering data for $\rho$ the countable set:
\begin{equation}\label{qsd}
\{k_n,c_n\}.
\end{equation}

The Camassa-Holm equation is obtained from 
the commutator relationship \reff{lax}, with $L$ given by \reff{chiso} and
$$
A=-\frac{\partial}{\partial t}+aD-\frac12 a_x, 
\qquad
a=u(x,t)+\frac{1}{k^2}.
$$
Note that the interval $(-M,M)$ remains constant, in view of \reff{interval}
and \reff{cholm2}.

To determine the evolution of the coupling coefficients, we proceed as in 
the Harry Dym flow, with suitable modifications. 
The compatibility relation \reff{lax} implies that the kernel of $L$ is 
invariant under $A$; hence $A\varphi(x,t,k)$ is a linear combination of 
$\varphi(x,t,k)$ and $\psi(x,t,k)$. The exact linear combination is determined 
by evaluating the asymptotic behavior of $A\varphi$ as $x\to\pm\infty$.
Under our assumptions, 
$u$ and $u_x$ tend to zero exponentially as $x$ tends to 
$-\infty$.  Again, the leading asymptotics of $\varphi$ are independent of
$t$, so we find that the 
leading asymptotics of $A\varphi$ as $x\to-\infty$ are
$$
A\varphi= (-\partial_t+aD-\frac12 a_x)\varphi\ \sim\ \frac{De^x}{k^2}=\frac{e^x}{k^2},
$$
hence
$$
A\varphi(x,t,k)=\frac{1}{k^2}\varphi(x,t,k).
$$
Similarly,
$$
A\psi(x,t,k)=-\frac{1}{k^2}\psi(x,t,k).
$$

At a bound state,
$$
\varphi(x,t,k_n)=c_n(t)\psi(x,t,k_n),
$$
hence
\begin{align*}
{c_n\over k_n^2}\,\psi_n&=\frac{1}{k_n^2}\,\varphi_n=A\varphi_n=A(c_n\psi_n)\\[4mm]
&= (-\dot c_n\psi_n+c_nA\psi_n)=
\left(-\dot c_n-\frac{c_n}{k_n^2}\right)\psi_n.
\end{align*}
Since $\psi_n$ is not identically zero, we have
\begin{equation}\label{evcoup}
\dot c_n(t)=-\frac{2c_n}{k_n^2},
\qquad
c_n(t)=e^{-2t/k_n^2}c_n(0).
\end{equation}

We next turn to the characterization of the image of 
$\rho^{-2}(D^2-1)$ under the Liouville transformation to
a Schr\"odinger operator. 

\medskip
\begin{theorem}\label{rangech} 
The necessary and sufficient conditions that the Schr\"odinger operator 
$D_\xi^2-q$ on the interval $(-M,M)$ be in the range of the Liouville 
transformation associated to \reff{chiso},
where $\rho=g(\tanh\,x)\text{sech}^2 x$ and  $g$ is a positive smooth function
on $[-1,1]$ with $g(-1)=g(1)$, are that $q$ be smooth on $[-M,M]$
and that the Dirichlet spectrum of $D_{\xi}^2-q$ be strictly negative.
\end{theorem}

\Proof  It is clear from the formula that describes $q$ in
terms of $g_0$ that $q$ must be smooth, and we have noted above that
the Dirichlet spectrum must be strictly negative.

Conversely, suppose that the Schr\"odinger potential $q$ is smooth 
and the Dirichlet spectrum is strictly negative.  By standard
oscillation theory the non-zero solutions to 
$$
(D_\xi^2-q)\varphi_0=0,\quad \varphi_0(-M)=0,\qquad (D_\xi^2-q)\psi_0=0, 
\quad\psi_0(M)=0
$$
do not change sign on the interval.  Therefore the (unique)
solution to the Dirichlet problem
$$
(D_\xi^2-q)\psi=0,\quad\psi(-M)=\psi(M)=1
$$
is a linear combination of $\varphi_0$ and $\psi_0$ that is
strictly positive on the closed interval.
The inverse Liouville transformation must be associated to the
change of variables $d\zeta/d\xi=\lambda/\psi(\xi)^2$,
where $\lambda$ is determined by the normalization condition
$$
\lambda\int^M_{-M}{d\xi\over \psi(\xi)^2}=2.\quad \qed
$$

\end{document}